\documentclass[aps,prl,twocolumn,floatfix,showpacs,superscriptaddress]{revtex4-1}
\usepackage{dcolumn}
\usepackage{graphicx}
\usepackage{amsmath}
\usepackage[colorlinks]{hyperref}

\begin{document}


\title{Orbital-Selective Mott Transition and Evolution of the Zhang-Rice State in Cubic Phase UO$_{2}$ Under Pressure}
\author{Li Huang}
\affiliation{Science and Technology on Surface Physics and Chemistry Laboratory, P.O. Box 718-35, Mianyang 621907, China}
\affiliation{Department of Physics, University of Fribourg, 1700 Fribourg, Switzerland}
\author{Yilin Wang}
\affiliation{Beijing National Laboratory for Condensed Matter Physics, and Institute of Physics, Chinese Academy of Sciences, Beijing 100190, China}
\author{Philipp Werner}
\affiliation{Department of Physics, University of Fribourg, 1700 Fribourg, Switzerland}
\date{\today}


\begin{abstract}
We study the electronic structure of cubic phase uranium dioxide at different volumes using a combination of density functional theory and dynamical mean-field theory. The \emph{ab initio} calculations predict an orbital-selective Mott insulator-metal transition at a moderate pressure of $\approx 45$ GPa. At this pressure the $j=5/2$ states become metallic, while the $j=7/2$ states remain insulating up to about 60 GPa. In the metallic state, we observe a rapid decrease of the 5$f$ occupation and total angular momentum with pressure. Simultaneously, the generalized Zhang-Rice state, which is of predominantly $j=5/2$ character, quickly disappears after the transition into the metallic phase. 
\end{abstract}


\maketitle


Over the past decades, actinide materials including pure elements, hydrides, oxides, carbides, and nitrides have been extensively studied with numerous experimental and theoretical tools, due to their fundamental importance in the nuclear energy industry and military technology~\cite{RevModPhys.81.235,actinidebook}. Even though great progress has been made, many problems and puzzles still remain. Of particular interest is the electronic structure of these actinide materials under extreme conditions (for example high pressure and high temperature in a reactor environment or a reactor accident)~\cite{Petit2014313,Benedict19921}.


Experimentally, it has been observed that under pressure many actinide materials undergo a series of structural phase transitions, such as the cubic to orthorhombic phase transitions occurring in some actinide dioxides~\cite{PhysRevB.70.014113}, or the three successive phase transitions in Am~\cite{PhysRevLett.85.2961}. To explain these complex phase transitions requires an accurate description of the properties of the $5f$ electrons over a wide range of pressures. It is well known that the $5f$ electrons, which play a pivotal role in determining the key physical and chemical properties of the actinide materials, react sensitively to changes in the surrounding environment~\cite{RevModPhys.81.235,actinidebook}. So, it is natural to expect that the 5$f$ electronic structure of actinide materials will be modified and some exotic correlation phenomena may emerge when an external pressure is applied. Up to now, only a few experiments and calculations have been conducted to explore the high pressure properties of actinide materials, and most of these efforts were devoted to study their structural phase transitions and instabilities~\cite{Benedict19921,PhysRevB.70.014113,PhysRevLett.85.2961,jpcm2003}. As a consequence, the high pressure electronic structures of actinide materials are poorly understood. Here, we employ a state-of-the-art first-principles approach to shed new light onto the pressure-driven electronic transitions in uranium dioxide. 


Stoichiometric UO$_{2}$ is in a cubic fluorite structure at ambient pressure. A transition to the orthorhombic cotunnite structure in the range of 42 $\sim$ 69\ GPa~\cite{PhysRevB.70.014113} was determined by high precision X-ray diffraction experiments. The ground state of UO$_{2}$ is an antiferromagnetic Mott insulator with a sizable band gap of $\approx$ 2.1\ eV~\cite{Baer1980885}. To study the electronic structure of UO$_{2}$, many traditional first-principles approaches have been employed, such as the local density approximation (LDA) [or generalized gradient approximation (GGA)] plus Hubbard $U$ approach~\cite{PhysRevB.75.054111,PhysRevB.88.024109,PhysRevB.79.235125,PhysRevB.88.104107,jpcm2009}, the hybrid functional method~\cite{wen2013}, and the self-interaction corrected local spin-density approximation (SIC-LSDA)~\cite{PhysRevB.81.045108}. However, none of the above methods can provide a satisfactory description of UO$_{2}$ over a wide range of conditions. For instance, though the band gap at ambient pressure is reproduced correctly, the phase transition pressures predicted by the LDA + $U$ and GGA + $U$ methods are 7.8~\cite{PhysRevB.75.054111} and 20 GPa~\cite{jpcm2009}, respectively, and thus much lower than the experimental value. This discrepancy is likely due to the incorrect treatment of the multiple metastable states of UO$_{2}$~\cite{PhysRevB.79.235125}. In addition, the changes of the electronic structure, especially the evolution of the Mott gap of UO$_{2}$ under pressure, is still an open and debated question~\cite{Petit2014313}.


In this letter, we use the density functional theory plus dynamical mean-field theory (DFT + DMFT) method~\cite{RevModPhys.68.13,RevModPhys.78.865}, which is currently the most powerful established tool to study strongly correlated systems, including the actinide materials~\cite{PhysRevB.84.195111,uru2si2,shim2007,zhu2013}, to calculate the electronic structure of cubic phase UO$_{2}$ under different pressures~\cite{details}. We predict that an orbital-selective Mott insulator-metal transition will happen around 45\ GPa. We also find that the pressure effect will lead to not only a redistribution of the 5$f$ electrons and an associated collapse of total angular momentum, but also a disappearance of the generalized Zhang-Rice state (ZRS). 


\begin{table}[t]
\caption{Calculated bulk properties and band gaps of cubic UO$_{2}$. Here $a_0$ denotes the lattice constant, $B$ the bulk modulus, and $B^{'}$ the first pressure derivative of $B$.\label{tab:bulk}}
\begin{ruledtabular}
\begin{tabular}{lllll}
                    & $a_0$ (\AA)                     & $B$ (GPa)                       & $B^{'}$ (GPa)                  & gap (eV) \\
\hline
The present work    & 5.571                           & 183.8                           & 4.24                           & 2.1 \\
LDA (GGA) + $U$     & 5.540~\cite{PhysRevB.88.024109} & 191.6~\cite{PhysRevB.88.024109} & -                              & 2.2~\cite{PhysRevB.88.024109} \\ 
                    & 5.449~\cite{PhysRevB.88.104107} & 222.4~\cite{PhysRevB.88.104107} & -                              & 2.3~\cite{PhysRevB.88.104107} \\
Hybrid functional   & 5.463~\cite{wen2013}            & 218.0~\cite{wen2013}            & -                              & 2.4~\cite{wen2013} \\
SIC-LSDA            & 5.400~\cite{PhysRevB.81.045108} & 219.0~\cite{PhysRevB.81.045108} & -                              & 0.0~\cite{PhysRevB.81.045108} \\
Experiments         & 5.473~\cite{PhysRevB.70.014113} & 207.2~\cite{PhysRevB.70.014113} & 4.54~\cite{PhysRevB.70.014113} & 2.1~\cite{Baer1980885} \\
\end{tabular}
\end{ruledtabular}
\end{table}

\begin{figure*}[t]
\centering
\includegraphics[width=\textwidth]{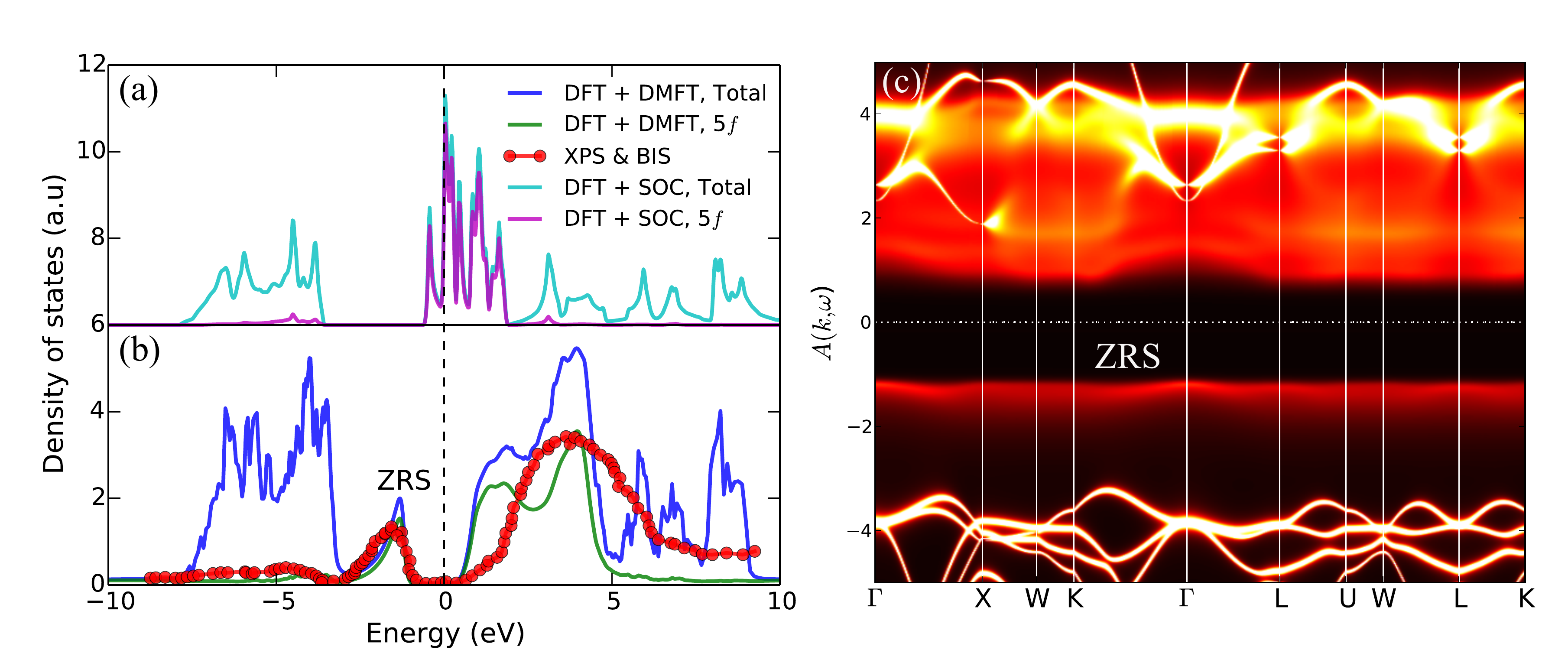}
\caption{(color online). Electronic structure of cubic UO$_{2}$ at ambient pressure. (a) Total and $5f$ partial density of states obtained by the DFT + SOC method. (b) Total and $5f$ partial density of states obtained by the DFT + DMFT method ($T\approx 116$ K). The XPS and BIS experimental data (denoted by red filled circles)~\cite{Baer1980885} are shown for comparison. (c) The momentum-resolved spectral function obtained by the DFT + DMFT method. Here ``ZRS" indicates the generalized Zhang-Rice state.\label{fig:gs_band}}
\end{figure*}

\emph{Physical properties at ambient pressure.} We first calculate the bulk properties of cubic phase UO$_{2}$ at ambient pressure and $T \approx$ 116\ K~\cite{details} to benchmark the accuracy of the DFT + DMFT approach. We use the equation of states (EOS) proposed by Teter \emph{et al.}~\cite{PhysRevB.52.8064} to fit the $E$-$V$ curve, and then extract useful parameters. The calculated bulk properties are summarized in Tab.~\ref{tab:bulk}. Other theoretical and experimental values, if available, are collected and listed as well. The obtained equilibrium lattice parameter $a_0$ is slightly overestimated, while the bulk modulus $B$ is  underestimated. The small deviations (1.7\% for $a_0$, 11.3\% for $B$, and 6.7\% for $B^{'}$) may be caused by the large Hubbard $U$ parameter used in the present calculations~\cite{details}. However, utilizing the obtained EOS, we find that at the experimental lattice volume, the calculated bulk modulus is 207.7\ GPa, which is almost identical with the experimental one~\cite{PhysRevB.70.014113}.

The electronic structure of UO$_{2}$ at ambient pressure is shown in Fig.~\ref{fig:gs_band}. We can see that the DFT + spin-orbit couping (SOC) method fails to reproduce the insulating nature of UO$_{2}$, and leads to a metallic state. Thus, considering the SOC effect alone is not sufficient for a proper description of the electronic structure of UO$_{2}$. In Fig.~\ref{fig:gs_band}(b) and (c), the total (partial) density of states and momentum-resolved spectral functions obtained by the DFT + DMFT approach are shown, together with the available XPS and BIS experimental data~\cite{Baer1980885}. From the calculated results, we can conclude that: (i) the energy gap is about 2.1\ eV, which is consistent with the experimental value (see Tab.~\ref{tab:bulk})~\cite{Baer1980885}. (ii) Since the energy gap is mainly associated with $5f \rightarrow 5f$ transitions, uranium dioxide is a Mott insulator, which agrees with the experiments~\cite{PhysRevLett.107.167406,PhysRevLett.106.207402} and most of the first-principles results~\cite{wen2013}. (iii) Between $-2.0$ and $-1.0$\ eV, there exists an isolated peak which is composed predominantly of U-$5f$ and O-$2p$ characters. The position and spectral weight of it agree quite well with the XPS experiments~\cite{Baer1980885,PhysRevLett.107.167406}. Recently, Yin \emph{et al.}~\cite{PhysRevB.84.195111} suggested that this low-energy resonance can be viewed as a generalized ZRS~\cite{PhysRevLett.100.066406}. Since the many-body ground state of the UO$_{2}$ $5f^2$ electronic configuration is a $\Gamma_5$ triplet, there must be a local magnetic moment which will couple with the O-$2p$'s hole induced by the photoemission process. Thus, in the following we identify this peak as the ZRS.


\begin{figure*}[t]
\centering
\includegraphics[width=\textwidth]{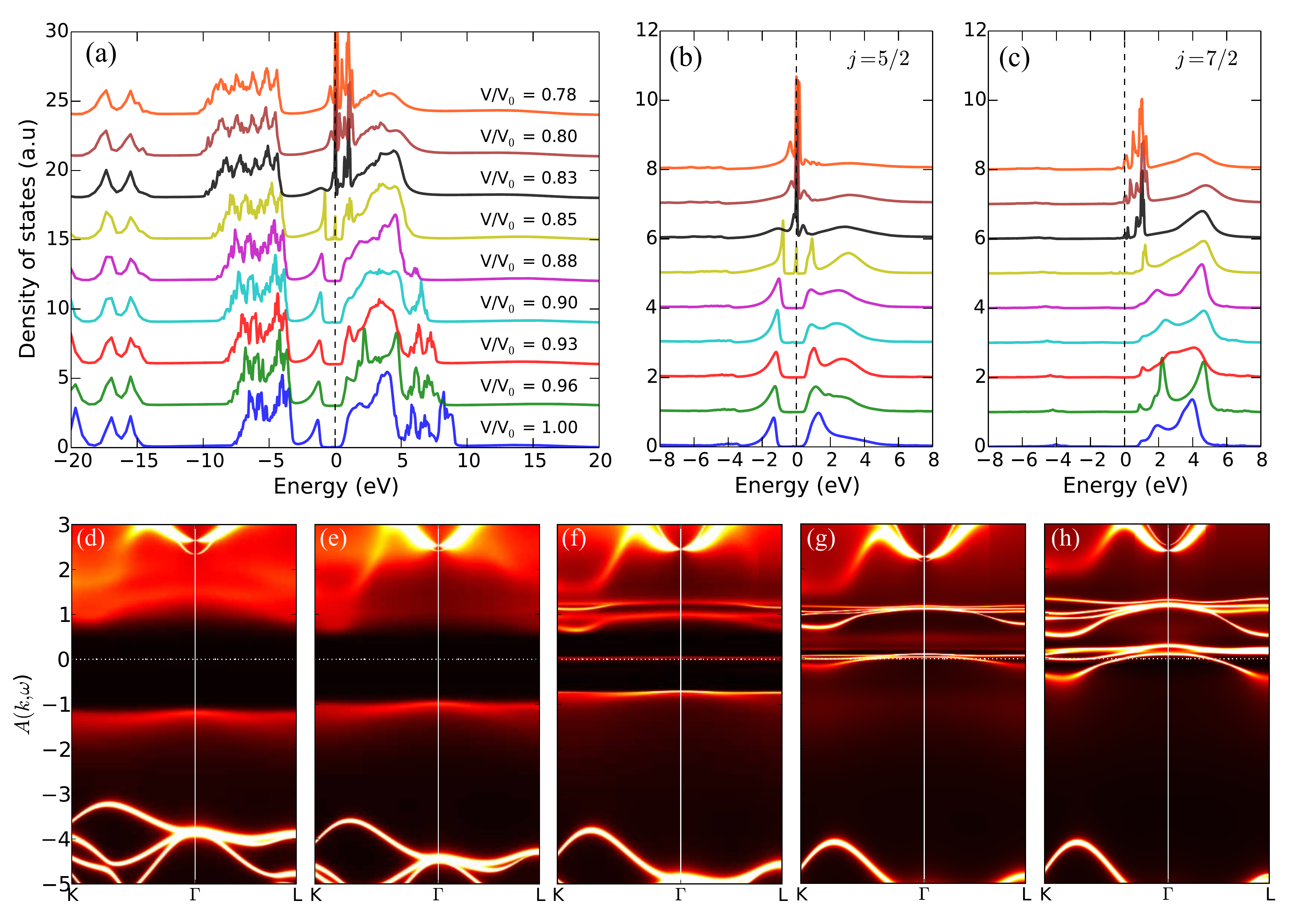}
\caption{(color online). Evolution of the electronic structure of cubic UO$_{2}$ under pressure calculated by the DFT + DMFT method. (a) Total density of states. (b) The U-$5f$ $j = 5/2$ partial density of states. (c) The U-$5f$ $j = 7/2$ partial density of states. (d)-(h) The momentum-resolved spectral functions for $V/V_{0}$ = 1.00, 0.90, 0.85, 0.83, and 0.78, respectively. Notice the sharp quasi-particle peak pinned at the Fermi level in panel (f).\label{fig:pressure_band}}
\end{figure*}

\begin{figure*}[t]
\centering
\includegraphics[width=\textwidth]{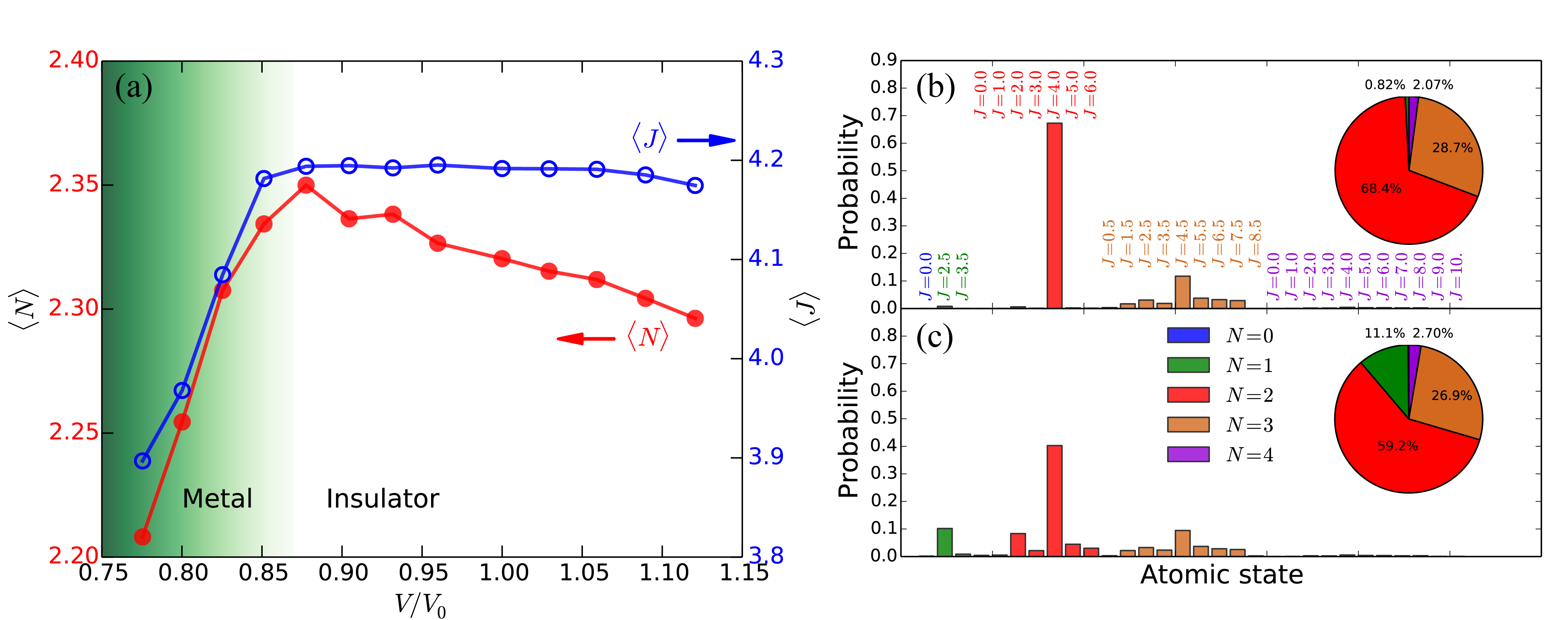}
\caption{(color online). (a) The average $5f$ electron occupancy $N_f = \langle N \rangle $ (left $y$-axis) and total angular momentum $\langle J \rangle$ (right $y$-axis) for various volumes. (b)-(c) Valence state histograms at $V/V_0$ = 1.00 (insulating region) and 0.78 (metallic region), respectively. The pie diagrams show the distributions of atomic states with respect to $N$. The atomic state probabilities for $N = 0$ and $N = 4$ states are too small to be visible.\label{fig:imp}} 
\end{figure*}

\emph{Mott insulator-metal transition.} Next, we concentrate on the pressure-driven electronic structure transition in cubic phase UO$_{2}$. We decrease the lattice constant $a_0$ to mimic the increase of external pressure, while the crystal structure (including the space group symmetry, which is Fm$\bar{3}$m) and the atomic coordinates are kept fixed. The evolution of the density of states with volume, as predicted by DFT + DMFT, is shown in Fig.~\ref{fig:pressure_band}(a). As the volume (pressure) is decreased (increased), dramatic changes are observed in the density of states. The band gap shrinks monotonously, and disappears suddenly at $V/V_{0}$ = 0.85, which marks a Mott insulator-metal transition. The critical transition pressure $P_c$ is about 45\ GPa according to the experimental $P$-$V$ curve~\cite{PhysRevB.70.014113}. When $P > P_c$, the quasi-particle peak grows quickly, which implies an enhancement of metallicity. The peak associated with the ZRS is also very sensitive to the volume collapse. Even though no strong shift of the ZRS peak to lower energies is apparent, which roughly agrees with previous DFT + DMFT calculations~\cite{PhysRevB.84.195111}, there is a substantial transfer of spectral weight to the quasi-particle peak. At $V/V_0$ = 0.80 (corresponding to $P \approx$ 65\ GPa)~\cite{PhysRevB.70.014113}, the ZRS peak is almost smeared out. The multiple peaks from $-12$\ eV to $-3$\ eV and from 1\ eV to 10\ eV are predominantly of O-$2p$ character~\cite{PhysRevLett.107.167406}. They are strongly hybridized with the lower and upper Hubbard bands of the U-$5f$ orbitals which are approximately located at $-4$ to $-5$ eV and $1$ to $6$ eV, respectively~\cite{PhysRevB.84.195111}. These ligand bands are also shifted outward and broaden under pressure. We further calculate the momentum-dependent spectral functions at various volumes in order to gain a better understanding of the evolution of the Mott gap and ZRS. Selected results are visualized in Fig.~\ref{fig:pressure_band}(d)-(h). These plots reveal the same Mott transition scenario as already shown in Fig.~\ref{fig:pressure_band}(a). From (d) to (f), the Mott gap is reduced from a finite value (2.1\ eV) to zero, and the ZRS feature is slightly shifted toward the Fermi level. In panel (f), where $V/V_0$ = 0.85, the quasi-particle peak appears, while the ZRS peak still persists. From (f) to (h), the ZRS peak is rapidly smeared out and at the same time, the weight of the quasi-particle peak is substantially enhanced.

Due to the SOC effect in heavy elements, the U-$5f$ orbitals are split into two components: $j = 5/2$ and $j = 7/2$ states (here we ignore the crystal field splitting for the sake of simplicity)~\cite{RevModPhys.81.235}. It is interesting to study the evolution of the corresponding density of states under pressure. The calculated results are shown in Fig.~\ref{fig:pressure_band}(b) and (c), respectively. If one compares the $j = 5/2$ partial density of states in Fig.~\ref{fig:pressure_band}(b) with the corresponding total density of states in Fig.~\ref{fig:pressure_band}(a), one can easily recognize that both the Mott gap and the ZRS peak are associated primarily with the $j = 5/2$ states. In Fig.~\ref{fig:pressure_band}(b), on can observe again how a quasi-particle peak grows at the Fermi level and the ZRS peak fades away when $V/V_0 \leq 0.85$. Thus, we propose that the $j = 5/2$ states are responsible for the pressure-driven Mott insulator-metal transition and the associated disappearance of the ZRS peak in cubic UO$_{2}$. As for the $j = 7/2$ states, most of the density of states is well above the Fermi level. Remarkably, even when $V/V_0$ = 0.85 (the approximate transition point for the $j = 5/2$ states), the spectral weight for the $j = 7/2$ states at the Fermi level remains zero. Only for $V/V_0 \leq $ 0.83, the $j = 7/2$ states become metallic. In other words, the Mott insulator-metal transitions for the $j = 5/2$ and $j = 7/2$ states don't occur simultaneously, and there is a sizable volume or pressure range ($ 0.83 \leq V/V_0 \leq 0.85$, corresponding to 45\ GPa $\leq P \leq$ 60\ GPa~\cite{PhysRevB.70.014113}) in which the $j = 5/2$ states are metallic while the $j=7/2$ states remain insulating. According to our calculations, the insulator-metal transition in cubic phase UO$_{2}$ is therefore an orbital-selective Mott transition~\cite{PhysRevLett.92.216402,PhysRevLett.99.126405}. 

Finally, we would like to emphasize that the structural phase transition is completely ignored in our calculations. In fact, the structural phase transition from cubic to orthorhombic begins at $V/V_0$ = 0.87, and continues beyond $V/V_0$ = 0.81. The transition zone extends at least up to 69\ GPa~\cite{PhysRevB.70.014113}. Since the Mott transition apparently occurs without any change in structure and the $P_c$ for it is very close to the one for the structural phase transition, we believe that the structural phase transition in UO$_{2}$ is driven by the Mott transition, which involves a localization-delocalization process of the U-$5f$ electrons. Previous theoretical research using the LDA + $U$ method predicted that the orthorhombic structure is an insulator, and the metallization pressure for it is in the range of 226 $\sim$ 294 GPa~\cite{PhysRevB.75.054111}. On the other hand, the SIC-LSDA approach predicts that the cubic UO$_{2}$ is on the verge of an insulator-metal transition at ambient pressure~\cite{Petit2014313}. Both predictions are inconsistent with ours and it would thus be very interesting to test them experimentally. 


\emph{5$f$ orbital occupancies and atomic multiplets.} The nominal 5$f$ occupation for UO$_{2}$ is 2. However, due to the strong hybridization between the U-$5f$ and O-2$p$ bands~\cite{PhysRevLett.107.167406}, the actual 5$f$ occupation is larger than 2. We analyzed the connection between the average 5$f$ orbital occupancy $N_f$ and the volume collapse $V/V_0$ [see Fig.~\ref{fig:imp}(a)]. Under compression, $N_f$ first increases steadily, and then decreases sharply. This behavior can be easily understood as follows: In the insulating phase, the U-$5f$ electrons are localized and the moderate changes in $N_f$ are due to the hybridization effect with the O-$2p$ bands. In the metallic phase, the U-5$f$ electrons become itinerant which allows them to hop from the $j = 5/2$ states to the $j = 7/2$ states and the O-$2p$ orbitals. As a result, one observes an abrupt decline of $N_f$ after the Mott insulator-metal transition.

The valence state histogram, which measures the probability to find a 5$f$ electron in a given atomic eigenstate, can provide additional information about the nature of the U-5$f$ electrons. Representative valence state histograms for $V/V_0$ = 1.00 (insulating phase) and 0.78 (metallic phase) are plotted in Fig.~\ref{fig:imp}(b) and (c), respectively. In the insulating phase, the dominated states have occupancy $N = 2$ and 3. The other states have only negligible weight. In the metallic phase, even though the $N = 2$ and 3 states remain dominant, the probabilities for the $N = 1$ states increase substantially. So, the change in the valence state histogram is consistent with the rapid decrease of $N_f$. We further calculated the effective total angular momentum $\langle J \rangle$ using $\langle J\rangle = \sum_{\Gamma} p_{\Gamma} J_{\Gamma}$ where $p_{\Gamma}$ is the probability for the atomic eigenstate $\Gamma$. In the insulating phase, $\langle J\rangle$ is hardly affected by the volume compression, while in the metallic phase it decreases steeply [see Fig.~\ref{fig:imp}(a)]. We note that the collapse of the total angular momentum under pressure is rather similar to the spin state crossover observed in some transition metal oxides~\cite{mno2008}, so it is likely the driving force behind the ``melting" of the ZRS peak~\cite{details}.


In summary, we predict that the Mott insulator-metal transition in cubic UO$_{2}$ occurs at $P_c \approx 45$\ GPa and is of the orbital-selective type~\cite{PhysRevLett.92.216402,PhysRevLett.99.126405}. There are two successive transitions for the $j = 5/2$ and $ j = 7/2$ states, respectively. The ZRS peak is only prominent in the insulating phase. In the metallic state, it quickly fades away with increasing pressure. At the same time, in the metallic phase, the pressure effect leads to a rapid decrease of the $5f$ occupancies and a collapse of the total angular moment. To our knowledge, this is the first study of the high pressure electronic structure of actinide dioxide using a modern DFT + DMFT approach. Our findings imply that physical properties such as resistivity, optical conductivity, and the magnetic moment of UO$_{2}$ should change radically across the (electronically driven) structural phase transition. We further speculate that other actinide dioxides, such as PaO$_{2}$, NpO$_{2}$, PuO$_{2}$, AmO$_{2}$, and CmO$_{2}$ etc., should exhibit similar pressure-driven phenomena~\cite{wen2013}. Thus, more high pressure experiments and theoretical calculations are highly desired. 


\begin{acknowledgments}
LH acknowledges support from the Swiss National Science Foundation (Grant No. 200021\_140648). YLW is supported by the National Science Foundation of China and the 973 program of China (No. 2011CBA00108). Most of the DFT + DMFT calculations were performed on the UniFr cluster (in Fribourg University, Switzerland) and TianHe-1A (in the National Supercomputer Center in Tianjin, China).
\end{acknowledgments}


\bibliography{zrs}

\end{document}